\documentclass[aps,pra,preprint,groupedaddress,showkeys,floatfix]{revtex4}

\begin{document}
\author{Aurelia Cionga$^1$, Fritz Ehlotzky$^2$ and Gabriela Zloh$^1$}
\affiliation{$^1$Institute for Space Science, P.O. Box MG-23, R-76900 Bucharest, Romania\\
$^2$Institute for Theoretical Physics, University of Innsbruck\\
Technikerstrasse 25, A-6020 Innsbruck, Austria}
\title{Circular dichroism in free-free transitions of high energy electron-atom
scattering}
\date{June 26, 2000}

\begin{abstract}
We consider high energy electron scattering by hydrogen atoms in the
presence of a laser field of moderate power and higher frequencies. If the
field is a superposition of a linearly and a circularly polarized laser beam
in a particular configuration, then we can show that circular dichroism in
two photon transitions can be observed not only for the differential but
also for the integrated cross sections, provided the laser-dressing of the
atomic target is treated in second order perturbation theory and the
coupling between hydrogenic bound and continuum states is involved.

\end{abstract}
\maketitle
This is the RevTeX shell.

\section{Introduction}

Dichroism is a well known concept in classical optics where it denotes the
property shown by certain materials of having absorption coefficients which
depend on the state of polarization of the incident light \cite{born}. This
concept has been further extended to the case of atomic or molecular
interactions with a radiation field. In particular, the notion of circular
dichroism in angular distribution (CDAD) refers to the difference between
the differential cross sections (DCS) of laser assisted signals for {\it left%
} ($L$) and {\it right} ($R$) circularly polarized ($CP$) light \cite{man1}.

Here we investigate the effect of the photon state of polarization, {\it i.e.%
} of its helicity, in laser-assisted high energy electron-hydrogen
scattering. We show under what conditions CDAD is observable at high
scattering energies as a result of {\it target dressing} by the laser field.
We consider optical frequencies and moderate field intensities and apply a
hybrid calculational approach \cite{byr1}. The interaction between the
projectile and the field is treated exactly, while the interaction between
the atom and the field is treated in perturbation theory. First order Born
approximation is used to evaluate the scattering amplitude. We demonstrate
that CDAD is encountered, provided i) the electromagnetic field is a
superposition of two laser beams, one of which is linearly polarized ($LP$)
and the other is a $CP$ field, ii) second order dressing of the target by
the electromagnetic field is included. In addition, iii) the role of the
virtual transitions to the continuum is shown to be essential for the
observation of CDAD. Finally, we demonstrate that for a special
configuration not only CDAD but also CD for the integrated cross sections
can be observed. Atomic units are used.

\section{Theory}

We consider electron-hydrogen scattering in the presence of an
electromagnetic field that is a superposition of two laser beams. One beam
is $LP$, with polarization vector $\vec{e}$, while the other is $CP$ with
polarization vector $\vec{\varepsilon}$. The beams can have different
directions of propagation. For simplicity, we discuss the case where the two
beams have the same frequency $\omega $ and intensity $I$. In dipole
approximation the resulting field is
\begin{equation}
\vec{{\cal E}}\left( t\right) =i\frac{{\cal E}_0}2\left( \vec{e}+\vec{%
\varepsilon}\right) \exp \left( -i\omega t\right) +{\rm c.c.},  \label{field}
\end{equation}
where the intensity $I={\cal E}_0^2$. We want to know whether the DCS are
sensitive to the helicity of the $CP$ photons, defined by
\begin{equation}
\xi =i\vec{n}\cdot \left( \vec{\varepsilon}\times {\vec{\varepsilon}}%
^{\;*}\right) ,  \label{hel}
\end{equation}
which explicitly depends on the direction $\vec{n}$ of propagation of the $%
CP $ beam. As shown in \cite{man1}, \cite{mitt} and \cite{inn1}, for high
energies of the projectiles CDAD does not occur for a $CP$ laser field {\it %
alone,} since the first order Born approximation leads to real scattering
amplitudes. We therefore present the theory for the above superposition of
fields.

According to \cite{byr1}, at moderate laser field intensities the field-atom
interaction can be described by time-dependent perturbation theory (TDPT).
We consider {\it second order} {\it dressing} of the hydrogen ground state
by the field (\ref{field}). The approximate solution for an atomic electron
in an electromagnetic field reads
\begin{equation}
|\Psi _1\left( t\right) >=e^{-i{\rm E}_1t}\left[ |\psi _{1s}>+|\psi
_{1s}^{(1)}>+|\psi _{1s}^{(2)}>\right] ,  \label{fun}
\end{equation}
where $|\psi _{1s}>$ is the unperturbed ground state of hydrogen, of energy $%
{\rm E}_1$. $|\psi _{1s}^{(1),(2)}>$ denote first and second order
corrections, respectively. On account of \cite{fhm} and \cite{f-m} these
corrections can be expressed in terms of
\begin{equation}
|{\vec{w}}_{1s}(\Omega )>=-G_C(\Omega )\vec{P}|\psi _{1s}>,  \label{lin}
\end{equation}
and
\begin{equation}
|w_{ij,1s}(\Omega ^{\prime },\Omega )>=G_C(\Omega ^{\prime })P_iG_C(\Omega
)P_j|\psi _{1s}>,  \label{sec}
\end{equation}
where $G_C({\Omega })$ is the Coulomb Green's function and $\vec{P}$ the
momentum operator of the bound electron. For the field (\ref{field}) there
are five values of the argument of the Green's functions necessary in order
to write down the approximate solution (\ref{fun}), namely $\Omega ^{\pm }=%
{\rm E}_1\pm \omega $, $\Omega ^{^{\prime }\pm }={\rm E}_1\pm 2\omega $, $%
\widetilde{\Omega }={\rm E}_1$.

A projectile of kinetic energy $E_k$ and momentum $\vec{k}$, moving in the
field (\ref{field}), is described by the Volkov solution
\begin{equation}
\chi _{\vec{k}}(\vec{r},t)=\frac 1{(2\pi )^{3/2}}\exp {\left\{ -iE_kt+i\vec{k%
}\cdot \vec{r}-i\vec{k}\cdot \vec{\alpha}(t)\right\} }.  \label{vol}
\end{equation}
$\vec{\alpha}\left( t\right) $ represents the classical oscillation of the
electron in the field ${\vec{{\cal E}}}(t)$, its amplitude is $\alpha _0=%
\sqrt{I}/\omega ^2$. Using Graf's addition theorem \cite{Wat}, the Fourier
expansion of (\ref{vol}) yields a series in terms of ordinary Bessel
functions $J_N$
\begin{eqnarray}
e^{-i\vec{k}\cdot {\vec{\alpha}}(t)} &=&\exp \left\{ -i\alpha _0\vec{k}\cdot
\vec{e}\sin \omega t-i{\cal R}_k\sin \left( \omega t-\phi _k\right) \right\}
\nonumber \\
&=&\sum_NJ_N({\cal Z}_k)\exp (-iN\omega t)\exp (iN\psi _k).  \label{dev1}
\end{eqnarray}
According to the definitions of the arguments and phases given in Watson's
book \cite{Wat}, we have
\begin{equation}
{\cal Z}_k=\alpha _0|\vec{k}\cdot \vec{e}+\vec{k}\cdot \vec{\varepsilon}%
\;|,\;{\cal R}_k=\alpha _0|\vec{k}\cdot \vec{\varepsilon}\;|,  \label{argc}
\end{equation}
and
\begin{equation}
\exp (i\psi _k)=\frac{\vec{k}\cdot \vec{e}+\vec{k}\cdot \vec{\varepsilon}}{|%
\vec{k}\cdot \vec{e}+\vec{k}\cdot \vec{\varepsilon}|},\;\exp (i\phi _k)=%
\frac{\vec{k}\cdot \vec{\varepsilon}}{|\vec{k}\cdot \vec{\varepsilon}|}.
\label{exp1}
\end{equation}
${\cal R}_k$ and $\phi _k$ refer to the $CP$ field alone, while ${\cal Z}_k$
and $\psi _k$ are related to the superposition (\ref{field}). Using (\ref
{exp1}), we recognize that a change of helicity of the $CP$ photons, {\it %
i.e.} $\vec{\varepsilon}\to \vec{\varepsilon}^{\;*}$, leads to a change in
sign of the dynamical phases $\phi _k$ and $\psi _k$. Therefore, looking for
the signature of helicity in the angular distributions of laser-assisted
signals, we have to observe the presence of these dynamical phases in their
DCS.

For high scattering energies, the first order Born approximation in terms of
the interaction potential is reliable. Neglecting exchange effects, this
potential is $V(r,R)=-1/r+1/|\vec{r}+\vec{R}|$, and the $S-$matrix element
reads
\begin{equation}
S_{if}^{B1}=-i\int_{-\infty }^{+\infty }dt<{\chi }_{{\vec{k}}_f}(t)\Psi
_1(t)|V|{\chi }_{{\vec{k}}_i}(t)\Psi _1(t)>,  \label{GZ}
\end{equation}
where $\Psi _1$ and ${\chi }_{{\vec{k}}_{i,f}}$ are given by (\ref{fun}) and
(\ref{vol}). $\vec{k}_{i(f)}$ are the initial(final) electron momenta.

The DCS for a process in which $N$ photons are involved is
\begin{equation}
\frac{d\sigma _N}{d\Omega }={(2\pi )}^4\frac{k_f{(N)}}{k_i}|T_N|^2.
\label{sed}
\end{equation}
The scattered electrons have the final energy $E_f=E_i+N\omega $ where $N$
is the net number of photons exchanged between the colliding system and the
field (\ref{field}). $N\geq 1$ refers to absorption, $N\leq -1$ to emission
and $N=0$ describes the elastic process. The nonlinear transition matrix
elements $T_N$ in (\ref{sed}) have the general structure
\begin{equation}
T_N=\exp \left( iN\psi _q\right) \left[ T_N^{(0)}+T_N^{(1)}+T_N^{(2)}\right]
.  \label{tm-cp}
\end{equation}
$\psi _q$ is the dynamical phase in (\ref{exp1}) evaluated for the momentum
transfer $\vec{q}={\vec{k}}_i-{\vec{k}}_f$. The first term in (\ref{tm-cp}),
\begin{equation}
T_N^{(0)}=-\frac 1{4\pi ^2}f_{el}^{B1}J_N({\cal Z}_q)\;,  \label{t0}
\end{equation}
yields the Bunkin-Fedorov formula \cite{Bunk} (target dressing is
neglected). Here $T_N=\exp \left( iN\phi _q\right) T_N^{(0)}$ and the Bessel
function $J_N\left( {\cal Z}_q\right) $ contains all the field intensity
dependences of the transition matrix element. $f_{el}^{B1}$ is the amplitude
of elastic scattering in the first order Born approximation, $%
f_{el}^{B1}=2\left( q^2+8\right) \left( q^2+4\right) ^{-2}$.

The remaining terms in (\ref{tm-cp}) describe the dressing of the atom by
the field (\ref{field}), they were discussed in detail in \cite{inn1}. In
the case of $T_N^{(1)}$ {\it one} of the $N$ photons exchanged between the
colliding system and the field is interacting with the bound electron, while
in $T_N^{(2)}$ {\it two} of the $N$ photons interact with the atomic
electron. These dressing terms in (\ref{tm-cp}) are
\begin{eqnarray}
T_N^{(1)} &=&\frac{\alpha _0\omega }{4\pi ^2q^2}\frac{|\vec{q}\cdot \vec{e}+%
\vec{q}\cdot \vec{\varepsilon}|}q  \nonumber \\
&&\times [J_{N-1}({\cal Z}_q)\;-J_{N+1}({\cal Z}_q)]{\cal J}_{1,0,1}(\tau
^{+},\tau ^{-},q)  \label{t1}
\end{eqnarray}
and
\begin{equation}
\begin{array}{l}
T_N^{(2)}=\frac{\alpha _0^2\omega ^2}{8\pi ^2q^2} \\
\times \{J_{N-2}({\cal Z}_q)[{|\vec{q}\cdot \vec{e}+\vec{q}\cdot \vec{%
\varepsilon}|^2}{q^{-2}}{\cal T}_1+\left( 1+2\vec{e}\cdot {\vec{\varepsilon}}%
\right) e^{-2i\psi _q}{\cal T}_2] \\
+J_{N+2}({\cal Z}_q)[{|\vec{q}\cdot \vec{e}+\vec{q}\cdot \vec{\varepsilon}|^2%
}{q^{-2}}{\cal T}_1+(1+2\vec{e}\cdot {\vec{\varepsilon}}^{\;*})e^{2i\psi _q}%
{\cal T}_2] \\
+J_N({\cal Z}_q)[{|\vec{q}\cdot \vec{e}+\vec{q}\cdot \vec{\varepsilon}|^2}{%
q^{-2}}\widetilde{{\cal T}}_1+2(1+{\rm Re\;}\vec{\varepsilon}^{\;*}\cdot
\vec{e})\widetilde{{\cal T}}_2]\}.
\end{array}
\label{t2}
\end{equation}
The five radial integrals, ${\cal J}_{1,0,1}$, ${\cal T}_1$, ${\cal T}_2$, $%
\widetilde{{\cal T}}_1$ and $\widetilde{{\cal T}}_2$ in (\ref{t1})-(\ref{t2}%
), depend on $q=|\vec{q}\;|$ and on the parameters of the Coulomb Green's
functions. The integral ${\cal J}_{1,0,1}$ is a function of the two
parameters $\Omega ^{\pm }$ through $\tau ^{\pm }=1/\sqrt{-2\Omega ^{\pm }}$
(see \cite{ac1}), while ${\cal T}_1$, ${\cal T}_2$, $\widetilde{{\cal T}}_1$
and $\widetilde{{\cal T}}_2$ depend on four parameters \cite{inn1}. ${\cal T}%
_1$ and ${\cal T}_2$ are multiplied by $J_{N\mp 2}$ if both photons are
absorbed/emitted by the atomic electron. In the last line of (\ref{t2}), $%
\widetilde{{\cal T}}_1$ and $\widetilde{{\cal T}}_2$ are multiplied by $J_N$%
. Here two photons interact with the atomic electron, but one is emitted and
the other is absorbed. For our numerical calculations we used the analytic
expressions for the above five radial integrals presented in \cite{inn1} and
\cite{ac1}-\cite{ac2}. Equivalent expressions were published for the case of
single photon transitions in \cite{mq1}-\cite{joac} and for two photon
absorption/emission in \cite{br1}.

The transition matrix elements for first and second order dressing in (\ref
{t1})-(\ref{t2}) are written in a form that permits to analyze their
dependence on the dynamical phase $\psi _q$. We see that $T_N^{(0)}$ and $%
T_N^{(1)}$ do not depend on the helicity (\ref{hel}). On the other hand, $%
T_N^{(2)}$ has an explicit dependence on $\xi ,$ determined by the phase
factors that multiply ${\cal T}_2$ in (\ref{t2}). Due to the structure of $%
T_N^{(2)}$, it is evident that in the absence of the $LP$ component of the
field (\ref{field}) the dynamical phase $\psi _q$ is absent and hence there
is no CDAD. Indeed, both polarization terms multiplying ${\cal T}_2$ would
be zero since ${\vec{\varepsilon}}^{\;2}={\vec{\varepsilon}}^{\;*\;2}=0$.

This demonstrates the essential role of the $LP$ beam and the necessity to
include second order dressing of the target. In order to stress the
important role of the virtual transitions to the continuum, we shall
consider small scattering angles. Here the dressing of the target is
considerable and the CDAD effect can be large.

\section{Weak field limit}

For small arguments of the Bessel functions, {\it i.e}. either for weak
fields at any scattering angle or for moderate fields at small scattering
angles, we can keep the leading terms in (\ref{tm-cp}) only. We discuss in
some detail the case $N=2$. The corresponding matrix element is
\begin{equation}
T_2=\frac{\alpha _0^2}{8\pi ^2q^2}\left[ \left( \vec{q}\cdot \vec{e}+\vec{q}%
\cdot \vec{\varepsilon}\;\right) ^2{\cal A}+\left( 1+2\vec{e}\cdot {\vec{%
\varepsilon}}\right) {\cal B}\right] ,  \label{abs}
\end{equation}
where the amplitudes ${\cal A}$ and ${\cal B}$ depend on $q$ and on $\omega $
\begin{eqnarray}
{\cal A}\left( q;\omega \right) &=&-\frac{q^2}{2^2}\left[ f_{el}^{B1}-\frac{%
4\omega }{q^3}{\cal J}_{1,0,1}-\frac{4\omega ^2}{q^4}{\cal T}_1\right] ,
\label{a} \\
{\cal B}\left( q;\omega \right) &=&{\omega ^2}{\cal T}_2.  \label{b}
\end{eqnarray}

The DCS derived from (\ref{abs}) are
\begin{eqnarray}
\frac{d\sigma _2}{d\Omega } &=&\alpha _0^4\frac{k_f}{k_i}\frac 1{2^2q^4}
\nonumber \\
\times \{ &\mid &\vec{q}\cdot \vec{e}+\vec{q}\cdot \vec{\varepsilon}\mid
^4\mid {\cal A}\mid ^2+\mid 1+2\vec{e}\cdot \vec{\varepsilon}\mid ^2\mid
{\cal B}\mid ^2  \nonumber \\
&&+2{\rm Re}[(\vec{q}\cdot \vec{e}+\vec{q}\cdot \vec{\varepsilon})^2(1+2\vec{%
e}\cdot \vec{\varepsilon}^{\;*}){\cal A}{\cal B}^{\;*}]\}.  \label{fin}
\end{eqnarray}
They depend on the change of helicity only if ${\rm Im}\;{\cal A}\neq 0$ and
${\rm Im}\;{\cal B}\neq 0$. As shown in \cite{inn1} and \cite{ac1}, this is
true if virtual transitions to continuum states are energetically allowed,
{\it i.e}. if $\omega >|{\rm E_1|}$ or $2\omega >|{\rm E_1|}$.

CDAD, defined as the difference between the DCS for $LCP$ and $RCP$, follows
from (\ref{fin}) as
\begin{equation}
\Delta _C= -\frac{k_f}{k_i}\frac{\alpha _0^4}{q^4}\; {\rm Im} {\cal Q} \;
{\rm Im} \left( {\cal A}^{\;*} {\cal B}\right) ,  \label{cdad}
\end{equation}
where ${\cal Q}= (\vec e +\vec \varepsilon )^2 ( \vec q \cdot e + \vec q
\cdot \vec \varepsilon ^{\;*})^2$ and thus ${\rm Im} {\cal Q}$ gives the
angular depence of $\Delta_C$. We easily verify that the three conditions
i)-iii), quoted in the introduction, are all necessary to have $\Delta
_C\neq 0$, namely (i) ${\cal Q} =0$ if the $LP$ field is absent, (ii) ${\cal %
B}$ stems from second order target dressing, and (iii) ${\cal A}$ and ${\cal %
B}$ become real as soon as $2\omega <|{\rm E_1}|$.

Next we study the angular dependence of $\Delta _C$. Two cases are of major
interest:

$(I)$ If $\vec{e}\;||\vec{e}_{{\rm i}}$, then the superposition of $LP$ and $%
CP$ is equivalent to elliptic polarization ($EP$) and we write $\Delta
_C\equiv \Delta _E$. Hence
\begin{equation}
\Delta _E=\frac{k_f}{k_i}\frac{\alpha _0^4}{q^4}q_{{\rm i}}q_{{\rm j}}\left(
\sqrt{2}+1\right) ^2{\rm Im}\left( {\cal A}^{\;*}{\cal B}\right) ,
\label{edad}
\end{equation}
where $q_{{\rm i;j}}={\vec{e}}_{{\rm i;j}}\cdot {\vec{q}}$ are the
projections of ${\vec{q}}$ on the axes ${\vec{e}}_{{\rm i;j}}$ of the $CP$
vector $\vec{\varepsilon}=\left( \vec{e}_{{\rm i}}+i\vec{e}_{{\rm j}}\right)
/\sqrt{2}$. Here $\Delta_E$ leads to CDAD but integrating it over $\varphi $
in the azimuthal plane yields zero. Expression (\ref{edad}) is comparable to
that of elliptic dichroism in photoionization \cite{man2}.

$(II)$ If $\vec{e}\;||\vec{e}_{{\rm j}}$, then CDAD reads
\begin{equation}
\Delta _C=-\frac{k_f}{k_i}\frac{\alpha _0^4}{q^4}\left[ \frac{q_{{\rm i}}^2}{%
\sqrt{2}}+q_{{\rm i}}q_{{\rm j}}-\frac{q_{{\rm j}}^2}{\sqrt{2}}\right] {\rm %
Im}\left( {\cal A}^{\;*}{\cal B}\right) .  \label{cd}
\end{equation}
Taking in addition $\vec{e}\;||{\vec{k}}_i$, then the term proportional to $%
q_{{\rm j}}^2$ gets $\varphi $-independent and its contribution survives in
the azimuthally integrated cross sections. Thus, the choice of a privileged
direction in the problem, namely that of ${\vec{k}}_i$, introduces an
additional asymmetry. Therefore, the signature of the photon helicity
prevails not only in the angular distribution but also in the integrated
cross sections. To the best of our knowledge, this is the first case of a
laser-assisted process in which CD in the integrated cross sections is
encountered.

Finally, we stress the importance of the form of our $T$-matrix element (\ref
{abs}), since a more general structure
\begin{equation}
T_2=\left( \vec{\varepsilon}_1\cdot \vec{\varepsilon}_2\right) M+(\vec{%
\varepsilon}_1\cdot \vec{v}_2)(\vec{\varepsilon}_2\cdot \vec{v}_1)N
\label{pit}
\end{equation}
was also found in discussions of dichroism in other processes, like
two-photon ionization \cite{harm}-\cite{ago}, two-photon detachement of H$%
^{-}$ \cite{fire} or elastic X-ray scattering by ground state atoms \cite
{prat}. Of course, the meaning of the vectors $\vec{v}_{1;2}$ is specific to
each process. In all these cases dichroism is caused by interferences
between the real and imaginary parts of the amplitude and two terms with
different angular behavior are needed to get such interferences. In the
examples above, $M$ and $N$ were complex quantities. Our conditions, $\omega
>|{\rm E_1|}$ or $2\omega >|{\rm E_1|}$, serve the same purpose. Contrary to
our problem, in photoionization or photodetachment of unpolarized systems
there is no equivalent for our privileged direction $\stackrel{\rightarrow }{%
k}_i$ and hence only CDAD is observable.

\section{Results and discussion}

We present numerical results for CDAD and CD in laser-assisted
electron-hydrogen scattering at high energies. We analyze the DCS for $N=\pm
2$ since here the CDAD effects are large enough. Using the above formalism,
we evaluated the DCS (\ref{sed}) in the azimuthal plane as a function of $%
\varphi $ for a fixed scattering angle $\theta =20^{\circ }$ and for the
initial scattering energy $E_i=100$ eV. Our laser frequency was $\omega =10$
eV, taken close to an atomic resonance in order to enhance the CD effects,
and we chose the moderate field intensity $I=3.51\times 10^{12}$ Wcm$^{-2}$.
The initial electron momentum $\vec{k}_i$ was taken parallel to the $LP$
vector ${\vec{e}}$, both pointing along the $z-$axis and the $LP$ beam
propagated in the $(x,y)$-plane.

In case $(I)$ the $CP$ beam propagated in the $y-$direction and the
corresponding $CP$ vector $\vec{\varepsilon}=({\vec{e}}_z+i{\vec{e}}_x)/%
\sqrt{2}$ has helicity $\xi =1,$ known as $LCP,$ while $\vec{\varepsilon}%
^{\;*}$ has opposite helicity $\xi =-1$, representing $RCP$. In this
configuration we have $\vec{k}_i||\stackrel{\rightarrow }{e}||{\rm Re}(\vec{%
\varepsilon})$.

In Figure 1(a) we present for $N=2$ with $E_f=E_i+2\omega $ the $\varphi -$%
dependence of the DCS at $\theta =20^{\circ }$. The data for $LCP$ are shown
by a dotted line and for $RCP$ by a dashed line. Clearly, the laser-assisted
signals depend on the helicity of the photon. In Figure 1(b) we present the
results for CDAD (\ref{edad}). Since here $\Delta _E\sim \cos \varphi $, the
''$+$'' and ''$-$'' in the two lobes indicate this dependence. If $\Delta _E$
gets integrated over $\varphi $, the net CD effect is zero. Similar results
and conclusions are obtained for $N=-2$.

For case $(II)$ the $CP$ beam propagated in the $x-$direction and the $CP$
vector is $\vec{\varepsilon}=({\vec{e}}_y+i{\vec{e}}_z)/\sqrt{2}$ so that
now $\vec{k}_i||\vec{e}||{\rm Im}\left( \vec{\varepsilon}\right) .$ The $LP$
beam propagated as before.

Figure 2 shows the DCS (\ref{sed}) and the CDAD (\ref{cd}) for $N=2$ in
panel (a) and for $N=-2$ in panel (b). Dotted lines are for $LCP$ and dashed
ones for $RCP$. Full lines are used for $\Delta _C$ (\ref{cd}) where
explicitly
\begin{eqnarray}
q_{{\rm j}}^2-{\sqrt{2}}q_{{\rm i}}q_{{\rm j}}-q_{_{{\rm i}}}^2
&=&(k_i-k_f\cos \theta )^2-k_f^2\sin ^2\theta \sin ^2\varphi  \nonumber \\
&&+{\sqrt{2}}(k_i-k_f\cos \theta )k_f\sin \theta \sin \varphi .
\end{eqnarray}
Due to this angular dependence, the integration of\ $\Delta _C$ over $%
\varphi $ does not vanish. Since the final momentum $k_f$ depends on $N$,
the shape of the azimuthal dependence of $\Delta _C$ is different for
absorption and emission. Our data show that the maximum value of dichroism
can amount up to $2/3$ of the assisted signal. This is comparable to or even
larger than the effect predicted for X-ray scattering \cite{prat} or for
two-photon ionization \cite{ago}. Similar to the case of X-ray scattering,
the dichroism in our case is increasing with increasing laser frequency. In
free-free transitions at high scattering energy the dichroic effects stem
from target dressing, which is increasing with the photon frequency. Target
dressing is significant for rather small scattering angles. We therefore
expect that dichroism is large in this angular domain, and not near $\theta
=\pi /2$ as for X-ray scattering and two-photon ionization.

\section{Summary and conclusions}

Summarizing, we considered scattering of high energy electrons by hydrogen
atoms in the presence of a laser field of moderate power but higher
frequencies. The field had two components of equal frequency and intensity.
One of the components was circularly, the other linearly polarized. The two
laser beams were permitted to propagate in different directions. In the
first order Born approximation we showed that in the above scattering
configuration CDAD becomes observable in two-photon transitions if
laser-dressing of the atomic target is carried out in second order TDPT and
transitions between atomic bound and continuum states are energetically
allowed, requiring higher laser frequencies. Since the scattering
probabilities decrease with increasing $\omega $, we prefered to choose $%
\omega $ close to an atomic resonance to enhence the signals. thanks to the
above laser configuration and second order target dressing, the $T$-matrix
elements become complex as a prerequisite for predicting CDAD\ in the Born
approximation, since the elastic Born-amplitude is real.

We conclude that at high projectile energies CDAD in free-free transitions
is a second order field-assisted effect occurring under special conditions
only. In particular, the matrix elements of the process considered have to
be complex. Similar effects might occur, if higher order terms of the Born
series are taken into account, since we know that then the scattering matrix
elements become complex.

Finally, we stress the role of the two asymmetries which were introduced
into the scattering configuration in order to obtain helicity dependent
nonlinear signals. One asymmetry came in by the $LP$ laser. This was
sufficient to achieve CDAD. A second asymmetry was determined by the
momentum $\vec{k}_i$ of the ingoing electrons. The two asymmetries together,
more precisely $\vec{k}_i||\vec{e}||{\rm Im}\left( \vec{\varepsilon}\right) $%
, then led to CD that even persists if the DCS are integrated over $\varphi $%
. In photoionization and photodetachement there is no equivalent to $\vec{k}%
_i$. Hence, in those processes only CDAD is encountered. The analysis of the
structure of the two photon transition matrix element led us to a general
formula for CDAD and to our undestanding of the physical reasons of the
configuration necessary to obtain CDAD and CD, respectively.\vspace*{0.5cm}

\acknowledgments{
We acknowledge a most valuable and enlightening discussion on circular
dichroism with Professor Anthony F. Starace. One of us (AC) wants to thank
Professor P. Agostini for his informations on preliminary results on CDAD in
photoionization in a two-colour laser field. This work was supported by the
Jubilee Foundation of the Austrian National Bank, project number 6211, and
by a special research project for 2000/1 of the Austrian Ministry of
Education, Science and Culture.}

\newpage
{\large Figure Captions}\newline
Fig. 1: Refers to case $(I)$, {\it i.e.} $\vec{k}_i||\stackrel{\rightarrow }{%
e}||{\rm Re}(\vec{\varepsilon})$. We present for $N=2$ the DCS as function
of the angle $\varphi $ at the scattering angle $\theta =20^{\circ }$. The
initial electron energy is $E_i=100$ eV, the radiation frequency is $\omega
=10$ eV and its intensity $I=3.51\times 10^{12}$ Wcm$^{-2}$. The panel (a)
shows the data for {\it LCP }as dotted line and the data for {\it RCP }as
dashed line. Clearly, the laser-assisted signals depend on the helicity of
the photons. In panel (b) the CDAD effect is visible but integration over $%
\varphi $ yields zero.\newline
Fig. 2: Treats case $(II)$, {\it i.e.} $\vec{k}_i||\stackrel{\rightarrow }{e}%
||{\rm Im}(\vec{\varepsilon})$. For the same parameter values as in Fig. 1,
we show the $\varphi $-dependence of the DCS in panel (a) for $N=2$ and (b)
for $N=-2$. Signals for {\it LCP} are dotted lines and signals for {\it RCP }%
are dashed lines. The CDAD effects are represented by full lines.
Integrating these data over $\varphi ,$ a non-vanishing CD effect remains.
The dependence of the effects on photon emission/absorbtion is apparent.


\begin{thebibliography}{99}
\bibitem{born}  M. Born and E. Wolf, {\it Principles of Optics} (Pergamon
Press, 1991).

\bibitem{man1}  N. L. Manakov, S. I. Marmo, and V. V. Volovich, Phys. Lett.
A {\bf 204}, 48 (1995).

\bibitem{byr1}  F. W. Byron Jr. and C. J. Joachain, J. Phys. B {\bf 17},
L295 (1984).

\bibitem{mitt}  M. H. Mittleman, J. Phys. B {\bf 26}, 2709 (1993).

\bibitem{inn1}  A. Cionga, F. Ehlotzky, and G. Zloh, Phys. Rev. A {\bf 61}
063417 (2000).

\bibitem{fhm}  V. Florescu, A. Halasz, and M. Marinescu, Phys. Rev. A {\bf 47%
},394 (1993).

\bibitem{f-m}  V. Florescu and T. Marian, Phys. Rev. A {\bf 34}, 4641 (1986).

\bibitem{Wat}  G. N. Watson, {\it Theory of Bessel Functions}, 2nd. Ed.
(University Press, Cambridge, 1962), p. 359.

\bibitem{Bunk}  F. V. Bunkin and M. V. Fedorov, Zh. Eksp. Theor. Fiz. {\bf 49%
}, 1215 (1965) [{\it Sov. Phys. JETP}, {\bf 22}, 884 (1966)].

\bibitem{ac1}  A. Cionga and V. Florescu, Phys. Rev.A {\bf 45}, 5282 (1992).

\bibitem{ac2}  A. Cionga and G. Zloh, unpublished.

\bibitem{mq1}  A. Dubois, A. Maquet, and S. Jetzke, Phys. Rev. A {\bf 34},
1888 (1986).

\bibitem{mq2}  A. Dubois and A. Maquet, Phys. Rev. A {\bf 40}, 4288 (1989).

\bibitem{joac}  P. Francken and C. J. Joachain, Phys. Rev. A {\bf 35}, 1590
(1987).

\bibitem{br1}  G. Kracke, J. S. Briggs, A. Dubois, A. Maquet, and V.
Veniard, J. Phys. B {\bf 27}, 3241 (1994).

\bibitem{man2}  N. L. Manakov, A. Maquet, S. I. Marmo, V. V\'{e}niard, and
G. Ferrante, J. Phys. B {\bf 32}, 3747 (1999).

\bibitem{harm}  P. Lambropoulos and X. Tang, Phys. Rev. Lett. {\bf 61}, 2506
(1988); H. G. Muller, G. Petite, and P. Agostini, {\it ibid.} 2507 (1988).

\bibitem{eu}  A. Cionga, Rom. J. Phys. {\bf 28}, 483 (1993)

\bibitem{ago}  R. Taieb, V. V\'{e}niard, A. Maquet, N. L. Manakov, and S. I.
Marmo, {\bf 62}, 013402 (2000).

\bibitem{fire}  M. V. Frolov, N. L. Manakov, S. I. Marmo, and A. F. Starace,
at ICAP 2000, 4-9 June 2000, Florence.

\bibitem{prat}  N. L. Manakov, A. V. Meremianin, J. P. J. Carney, and R. H.
Pratt, Phys. Rev. A {\bf 61}, 032711 (2000).
\end{thebibliography}
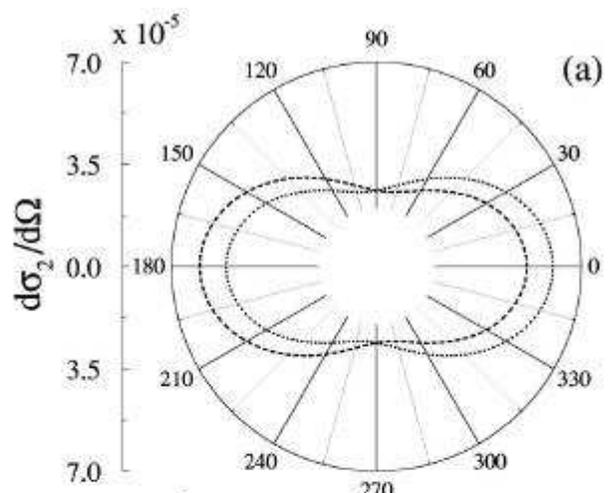
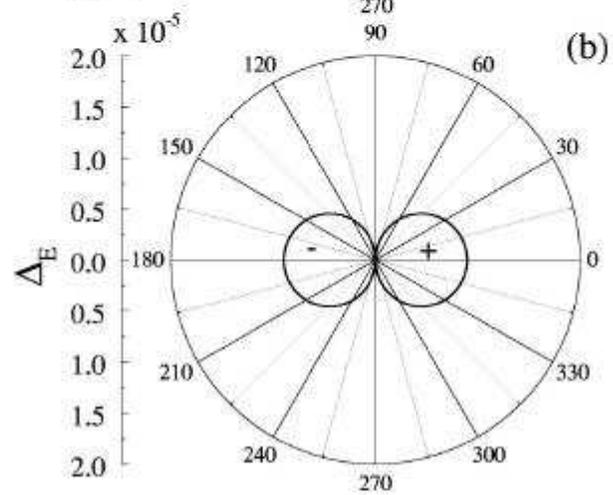
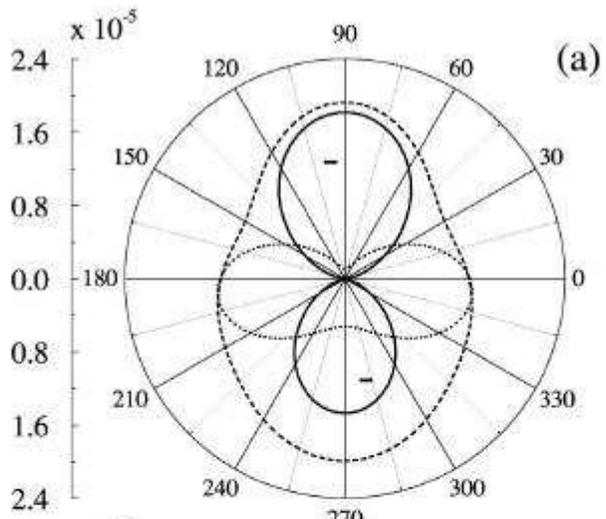
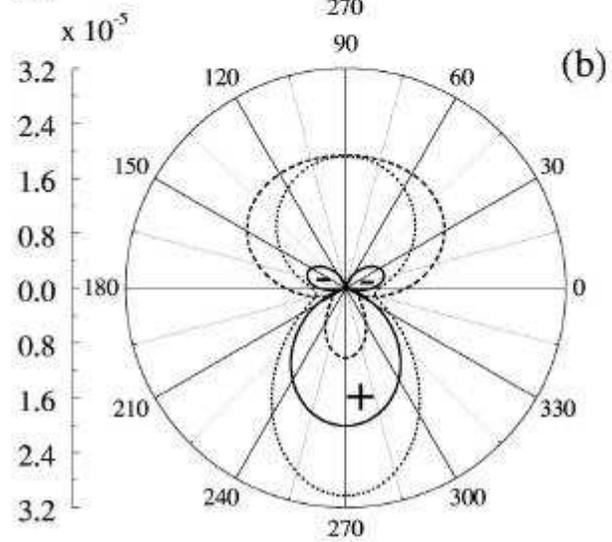
\end{document}